# A review of computational tools for generating metagenome-assembled genomes from metagenomic sequencing data

Chao Yang[1†], Debajyoti Chowdhury[2,3†], Zhenmiao Zhang[1], William K. Cheung[1], Aiping Lu[2,3], Zhao Xiang Bian[4,5], Lu Zhang[1,2*]

[1]Department of Computer Science, Faculty of Science, Hong Kong Baptist University, Hong Kong SAR, Hong Kong

[2]Computational Medicine Lab, Hong Kong Baptist University, Hong Kong SAR, Hong Kong

[3]Institute of Integrated Bioinformedicine and Translational Sciences, School of Chinese Medicine, Hong Kong Baptist University, Hong Kong SAR Hong Kong

[4] Institute of Brain and Gut Research, School of Chinese Medicine, Hong Kong Baptist University, Hong Kong SAR, China

[5] Chinese Medicine Clinical Study Center, School of Chinese Medicine, Hong Kong Baptist University, Hong Kong SAR, China

† These authors contributed equally to this work

*To whom correspondence should be addressed: E-mail: ericluzhang@hkbu.edu.hk

**Abstract:**

Microbes are essentially yet convolutedly linked with human lives on the earth. They critically interfere in different physiological processes and thus influence overall health status. Studying microbial species is used to be constrained to those that can be cultured in the lab. But it excluded a huge portion of the microbiome that could not survive on lab conditions. In the past few years, the culture-independent metagenomic sequencing enabled us to explore the complex microbial community coexisting within and on us. Metagenomics has equipped us with new avenues of investigating the microbiome, from studying a single species to a complex community in a dynamic ecosystem. Thus, identifying the involved microbes and their genomes becomes one of the core tasks in metagenomic sequencing. Metagenome-assembled genomes are groups of contigs with similar sequence characteristics from de novo assembly and could represent the microbial genomes from metagenomic sequencing. In this paper, we reviewed a spectrum of tools for producing and annotating metagenome-assembled genomes from metagenomic sequencing data and discussed their technical and biological perspectives.

# 1. Introduction

Microbes are essential for lives, and humans are intricately linked with microbial communities that are involved in different physiological processes[1,2]. Despite their crucial associations in physiology, the insights encompassing their co-existence were poorly characterized in the past. It was traditionally addressed using a culture-dependent way (Figure 1A) that enables isolating and sequencing individual microbes from the lab-based culture[3,4]. However, it is limited in identifying the broad spectrum of microbes and their co-evolution within hosts[5,6]. Despite having a new front of culturing efforts, a large yet undetermined range of microbial diversity has remained uncharacterized within the gut ecosystem[7–9]. This challenge is substantially addressed by the metagenomic approach (also known as the culture-independent method, Figure 1A) that studies the complex microbial communities in which they usually exist[3]. This technology enables to retrieval of genomes right from a mixture of microbial genomes in a culture-independent way without species isolation[9–11]. Subsequently, the next generation sequencing (NGS) technologies along with advanced computational tools have been extensively incorporated in analyzing and interpreting metagenomics data to investigate the diverse sphere of the microbiome[9,12–16].

Many investigations have adopted metagenomic sequencing to explore their impacts on human health. This has opened a new window in medical science and revealed a great extent of novel associations among the host microbiome and diseases. For instance, a meta-analysis study stated a link of several gut microorganisms, such as *Fusobacterium nucleatum*, *Parvimonas micra*, and *Gemella morbillorum* with colorectal cancer while studying several gut metagenomic sequencing datasets from colorectal cancer patients[17]. Another study by Thingholm LB, *et al.*, examined that several bacteria such as *Akkermansia*, *Faecalibacterium*, *Oscillibacter,* and *Alistipes* which are related to producing short-chain fatty acids, were significantly depleted in obese individuals and linked with their alterations of serum metabolites[18]. Later, the studies with the microbiome of the patients with Alzheimer's disease, a profound enrichment of bacteria was found to induce pro-inflammatory states, and it suggests a nexus between gut and brain[19,20]. Furthermore, the vaginal microbiome also gained rapid attention as their potential role in premature birth and Jennifer M. Fettweis *et al.* discovered some preterm-birth-associated taxa, such as *BVAB1*, *Sneathia amnii*, and *TM7-H1* among the most prominent candidates correlated with proinflammatory cytokines linked to the preterm-birth[21].

Most of those studies utilized the reference-based profiling strategy which directly aligns the short-reads against the reference genomes, marker gene sets or specific sequences for taxonomic assignments. The k-mer information or universal single-copy gene is usually extracted from the reference genomes and used to estimate relative sequence abundance and relative taxonomic abundance for interpreting data[22]. However, as the reference sequences are not complete, those aforesaid studies mostly rely on the well-studied genus or species. It creates a technical challenge to annotate novel genes, species, or strains. For example, approximately 40%-50% of human intestinal microbes lack a reference genome[23,24]. Therefore, it largely demands a well-characterized collection of reference genomes to annotate the abundance as well as the functions of novel microorganisms.

With recent advancements in metagenomics, those challenges are attempted to be addressed using the Metagenome-Assembled Genome (MAG) to represent the reference genome based on metagenome assembly, which enables the high throughput retrieval of microbial genomes right from samples in a culture-independent way without species isolation[9–11]. A MAG is sequence fragments asserted to be a close representation or an actual bacteria genome. In this approach, sequencing reads are assembled into contigs and then the contigs are binned into candidate MAGs based on their sequence context, abundance, and co-variation of abundance across the groups of samples[25,26]. Successively, these MAGs are subjected to quality checks and used for gene prediction and annotation further (Figure 1B). This approach has provided valuable bespoke reference databases for well-studied environments such as the human gut. For instance, through a large-scale assembly of metagenomic data from the human gut, Pasolli E and colleagues uncovered >150,000 microbial genomes and more than 50% of belonging species were never described before[27]. This discovery has increased the average mapping ability of gut metagenomes sequencing reads from 67.76% to 87.51%. Almeida A *et al.* also reported that the extent of gut proteins they identified from assembled genomes are more than two-folds of the number of genes present in the current database[28]. These newly identified genomes and genes have been demonstrated to have distinct functional properties and to be associated with numerous diseases, which may potentially improvise the current predictive models[9,29].

Here, we have reviewed different tools and techniques used to construct MAGs. They are primarily categorized into two sections, first, the upstream analyses, and second, the downstream analyses. The upstream analyses to construct MAGs include metagenome assembly, quality control and contig binning those are discussed in section 2; and downstream analyses for annotating MAGs, including gene prediction algorithms, gene functional annotation tools, and taxonomic profilers for MAGs have been discussed in section 3; and then the potential challenges in data analysis and a few plausible strategies to address them have been discussed in section 4. This review has demonstrated the enormous potential of metagenomic approaches in investigating microbiome, including a handful of reasonable solutions to overcome the current challenges associated with the technical limitations in this field of metagenomics.

## 2. Tools for upstream analyses to construct MAGs

### 2.1. Tools for metagenome assembly

**Metagenome assemblers for short-read sequencing**

Traditionally, most metagenomic assemblers for short-reads were designed using overlap-layout consensus (OLC) approaches (Table 1). For example, Omega[30] stores the prefix and suffix sequences for each read within the hash tables, and then it is used to construct a bi-directed graph upon linking the reads with their overlaps. This graph is later simplified by removing transitive edges that identify the paths with minimum cost. Due to some of the inherent issues associated with OLC, it is difficult for Omega to handle large sequencing reads, and it is also unable to distinguish the chimeric contigs. Many other assemblers are designed based on the De Bruijn graph (dbg), which splits the reads into k-mers and can reduce the computer memory cost (Table 1). Of them, the MetaVelvet[31], a popular metagenome assembler that is constructed based on Velvet[32]. MetaVelvet constructs the dbg with Velvet and introduces partitions into it to further create subgraphs using coverage peaks of the nodes. The chimeric contigs and the contigs with repetitive sequences are identified and split using paired-end information and local differences in coverage. MetaVelvet-SL[33] improves the decision-making process to deconvolve the chimeric contigs using support vector machines. MetaVelvet-DL[34] constructs end-to-end deep learning models with convolutional neural networks (CNNs) and Long Short-Term Memory units. It has been proved to be more powerful to decipher the chimeric contigs when compared to MetaVelvet-SL. However, a common problem of dbg is the selection of k-mer size, and it lays a substantial impact on its ability to deal with the repetitive sequences and the uneven node coverage[34]. Then, to optimize the choice of k-mers, IDBA-UD[35] attempts to prune the graph iteratively and merges bubbles with increasing k-mer sizes. The k-mer size will be determined if a significantly different coverage of the graph is observed. Also, MEGAHIT[36,37] couples the process of selecting k-mer sizes with succinct dbg and shows a strong computational efficiency. metaSPAdes[38], a very popular metagenomic assembler that improves the SPAdes[39] by introducing a novel heuristic strategy to differentiate the inter-species repeats. It assumes an uneven coverage in the assembly graph and builds multiple dbgs with different k-mer sizes. The hypothetical k-mers are designed to identify the chimeric contigs. Another advanced tool, Ray Meta[40] implements a strategy to generate local coverage distribution for each seed path of the assembly graph. It can be deployed explicitly on the distributed computing system and enable metagenome assembly on the computer clusters without large memory machines.

**Metagenome assemblers for long-read sequencing**

Due to the lack of long-range genome connectedness, the assemblers designed for short-read are generally limited to deal with the intra- and inter-species repeats. The adventure of long-read sequencing platforms, such as linked-read sequencing, synthetic long reads (SLRs), long reads from Oxford Nanopore Technologies (ONT), and Pacific Biosciences (PacBio) have shown great promises by generating virtual or physical long reads. For the co-barcoded linked-reads produced by the 10x Genomics Chromium platform, Bishara *et al.* [41] developed Athena that improves metagenomic assemblies by considering the co-barcoded reads between contigs. It constructs the scaffold graph by linking the contigs from metaSPAdes based on the support of paired-end reads. The local assembly is performed by recruiting the co-barcoded linked-reads shared by the contig pairs connected in the scaffold graph. cloudSPAdes[42] builds upon the assembly graph of metaSPAdes and evaluates the similarity between the barcode sets of two edges, which measures the probability of them to be derived from the same genomic region. The edges with high similarity would be connected to simplify the graph. Nanoscope[43] integrates SOAPdenovo[44] and Celera[45] to assemble the short-reads and SLRs independently and merges their contigs with Minimus2[46].

For PacBio and ONT long-reads, the assembler can be categorized as pre-and post-assembly error correction. The pre-assembly error correction tools, such as Canu[47] and NECAT[48], correct the sequencing errors in the long-reads and then assemble these clean reads by the OLC approach. Rather than correcting the error-prone long reads, wtdbg2[49] enables the process of inexact sequence matches to build the consensus from the intermediate contigs. metaFlye[50] that has been built upon Flye[51] with dedicated features for metagenome assembly. It combines the long-reads into error-prone disjointigs and collapses the repetitive sequences into a repeat graph.

**Metagenome assemblers for hybrid assembly**

The short-reads and long-reads from PacBio and ONT sequencing are complementary to each other. Several algorithms have been developed to make better use of the high base quality of short-reads and long-range genome connectedness of PacBio and ONT long reads. metaSPAdes[38] considers the error-prone long reads like the "untrusted contigs" and applies them to thread the complex structure of the assembly graph from short-reads. DBG2OLC[52] aligns the contigs from short-reads assembly to the error-prone long reads and applies the OLC strategy to concatenate those long-reads into contigs. OPERA-MS[53] utilizes the long-reads with shallow coverage to link the contigs from short-reads into an assembly graph and then groups them into species-specific clusters. It designs a novel Bayesian clustering algorithm to produce strain-resolved assembly using both contig coverage and contiguity information from long-reads. Unicycler[54] was initially designed to assemble a single bacterial genome and later be applied to metagenome assembly[55]. It only uses long-reads to choose the paths for the ambiguous regions in the assembly graph and applied multiple sequence alignment to correct their sequencing errors.

## 2.2 Tools for assembly quality control

As shown in Table 2, many tools have been developed to evaluate the correctness and continuity of the contigs generated by metagenome assemblers. MetaQUAST[56] can rapidly calculate the basic statistics for the contigs/scaffolds such as assembly length, N50, contig length distribution. It also supports a reference-based assessment in the "meta" mode, that aligns the sequences against the reference genomes from publicly available databases, wherein the statistics such as reference genome coverage, NGA50 used to rely on the availability of those reference genomes. REAPR[57] identifies the misassemblies using the insert size constraints of paired-end reads and pinpoints the regions with stretched distances between the aligned paired-end reads or the read depths are uneven. VALET[58] performs contig binning before quality control to reduce the false positives and false negatives due to uneven read depth. Recently, DeepMAsED[59] has been developed to detect the misassembled contigs without the need for reference genomes using a deep learning model.

## 2.3 Tools for contig binning

Because most of the current assemblers are not expected to produce complete microbial genomes from metagenomic samples. Many binning tools have been developed to group the contigs into clusters to represent the whole genome of an organism (Table 3). The existing tools rely mainly on the information of sequence composition such as tetranucleotide frequencies (TNF), k-mer frequencies, and read depth. MaxBin2[60] aims to recover the genomes from multi-sample co-assembly and jointly considers TNF and read depth using an expectation-maximization algorithm to calculate the distances between contigs. CONCOCT[25] extracts k-mer frequencies and combines them with the read depths to cluster contigs using Gaussian mixture models. MetaBAT2[61] also uses the above two features to calculate the contig similarities and constructs a graph by representing the similarities as the edges' weights. The graph is further partitioned into subgraphs (bins) based on a modified label propagation algorithm. Besides TNF and read depths, other information has also been taken into consideration in linking contigs. MyCC[62] implements an affinity propagation algorithm to make use of the complementary marker genes between contigs. SolidBin[63] performs a spectral clustering on taxonomy alignments to connect contigs; BMC3C[64] applies ensemble clustering on codon usage inferred from the contigs. Mallawaarachchi *et al.* identified the short contigs commonly neglected in the previous contig binning tools and developed GraphBin[65], which can label the short contigs by label propagation on the assembly graph. METAMVGL[66] integrates the assembly and paired-end graphs and develops a multi-view label propagation algorithm to correct the binning errors due to graph dead ends. VAMB[67] represents the contigs in low dimensions by using a deep learning-based autoencoder model. Despite many tools that have been developed for contig binning, there is no single best choice and the ensemble-based tools, MetaWRAP[68] and DAStool[69], have shown a promising way.

Besides short-read sequencing, Hi-C is another sequencing technology that has been applied for contig binning by introducing the genome's spatial proximity. ProxiMeta[70] devolves the plasmid genomes and generates the high-quality microbial bins without relying on prior information. Moreover, bin3C[71] designs an effective pipeline to generate contact map, bias removal, interaction strength normalization and applies the Louvain method[12] for contig community detection. HiCBin[72] adopts HiCzin[73] to normalize the interaction map and applies the Leiden community detection algorithm to group contigs. It also includes the module for spurious contact detection in the binning pipeline.

## 2.4 Decision making of MAGs

For each contig bin, the quality is commonly determined by the completeness of marker genes and the contamination of single-copy genes using CheckM[74]. Only the bins with relatively high quality would be selected as the MAGs for subsequent annotation. The contig bins are commonly classified as finished, high-quality, medium-quality, and low-quality drafts concerning completeness, contamination, and rRNA/tRNA prediction[75]. Due to the known issues in assembling rRNA/tRNA prediction[75],[76], it is well accepted to select high-quality (completeness >90% and contamination <5%) and medium-quality (completeness >50%, contamination <10% and completeness – (5 × contamination) >50) bins as MAGs[29].

## 3. Tools for downstream analyses to annotate MAGs

### 3.1. Gene prediction tools

Gene identification and annotation are the next essential steps after carefully selecting MAGs from metagenome assembly. However, it is often challenged by the short-reads that are difficult to be assembled. It makes the entire gene prediction procedure a challenging task. Here, we have discussed the approaches to identify and predict the genes to recognize biologically functional DNA sequences within genomes which play critical roles (Table 4).

**Model-based gene prediction tools**

The Hidden Markov model (HMM)-based tools are the most prevalent among the model-based gene prediction. There are several tools available under this category such as MetaGeneMark[78], Glimmer-MG[79], FragGeneScan[80]. MetaGeneMark extracts oligonucleotide frequencies and their compositions from the genomes of the known prokaryotic species to train the parameters in HMM model. Glimmer-MG clusters the input sequences that are most likely to share the same origin and trains the interpolated Markov model within each cluster to optimize the probabilistic models. FragGeneScan incorporates the sequencing error models into six-periodic inhomogeneous Markov models, enabling the identification of genes with frameshifts.

There are also several tools for gene prediction in bacterial and archaeal genomes based on dynamic programming. For example, Prodigal[81] applies dynamic programming on frame bias scores in the preliminary phase to train gene models. The same algorithm is adopted to process hexamer coding scores for each gene to produce the final gene list to predict the protein-encoding genes precisely. MetaGene[82] calculates two types of scores for all possible open reading frames (ORFs) measuring their intrinsic (including base composition and length) and extrinsic characters (including orientations and distance of neighbouring genes). These scores are further combined and served as input to the dynamic programming algorithm to further estimate the optimal path of ORFs. Later, MetaGeneAnnotator[83] improves the prediction ability of MetaGene.

**Deep learning-based gene prediction tools**

Recently, a variety of deep learning tools has gained considerable attention to augment gene prediction. Meta-MFDL[84] is a popular tool that constructs a representative vector by fusing multiple features like monocodon usage, mono amino acid usage, ORF length coverage, and Z-curve features and then trains a deep stacking network to classify the ORFs into coding or noncoding ORFs. CNN-MGP [85] automatically extracts the significant attributes from extracted ORFs and numerically adds them into a suitable convolutional neural network (CNN) relying on the GC content of input fragments. Balrog[86] aims to obtain the amino acid sequences from all available high-quality prokaryotic genomes and constructs a universal temporal CNN-based gene prediction tool for prokaryotic genomes as well as newly assembled genomes.

**3.2. Gene functional annotation tools**

Metagenomic sequencing enables the evaluation of the functional characteristics of microbial communities. The functional annotation tools for predicted genes can be classified into two categories, one, the tools with having broad scopes to evaluate the entire functional potential, and two, the tools with narrow scopes focusing on one or a few specific biological processes. This review will focus only on the tools designed with broad functional overviews (Table 5).

**Homology-based tools**

Traditionally, homology-based tools are used to employ different variants of BLAST[87] to compare the predicted genes with the sequences of those known genes. They are often very slow to process a huge number of genes predicted from MAGs. On the other hand, the modern methods employ optimized alignment strategies enabling 100 to 1000 folds more speed for aligning gene sequences to databases. The eggNOG-mapper[88], GhostKOALA[89], MG-RAST[90] and PANNZER2[91] are among them which are most commonly used. The eggNOG-mapper performs ultra-fast alignments of orthology relying on the pre-computed clusters and phylogenies information of the eggNOG[92] database. It uses HMM to search the best matching reference sequences for each query protein and performs faster than the traditional BLAST-based approach[88]. GhostKOALA[89] utilizes an automatic annotation approach relying on GHOSTX[93] supporting both genomic and metagenomic sequences and also assigns KEGG Orthology and pathways to each gene[94]. MG-RAST[90]provides an online metagenomic analysis interface, that includes data uploading, quality control, and alignment with M5nr reference databases[95], which includes UniProt[96], eggNOG[92], and KEGG[94]. PANNZER2[91] incorporates the SANSparallel[97], a MPI implementation of a suffix array neighbourhood search approach, to provide rapid homology searches for the Gene Ontology[98] annotations.

**Motif-based tools**

Sometimes the protein sequences are partially assembled, and it may cause adverse influence during homology-based annotation. It usually happens due to the incomplete and misassembled contigs of MAGs. In such cases, despite their poor alignment homologies, they tend to perform similar functions like those containing some common sequences, patterns, or specific motifs. Databases such as InterPro[99], PROSITE[100], and PRINTS[101] have collected such patterns or motifs based on statistical inference. InterProScan[102] uses Phobius[103] to process the proteins and domains from the InterPro database. Given an input sequence, it performs a systematic search within the InterPro database and predicts the protein domains, active sites, and potential functional annotations. However, as the extent of novelty within the MAGs is inherent, it is always recommended to perform both motif-based analysis and other homology-based approaches for better functional annotations.

**Gene context-based tools**

Metagenomic sequencing data enables the recognition of a large number of novel genes that may share no homology with the known genes and are not suitable to be annotated using those aforesaid approaches. To address such limitations, gene context-based tools have been introduced. Harrington *et al.* combines the homology-based approaches and tailored the gene neighbourhood methods to perform gene annotation for a complex metagenomic dataset[104]. It infers specific functions for 76% and non-specific functions for 83% of the sequences and outperforms the standard BLAST-based methods. Ciria R *et al.* develops GeConT [105] to illustrate the genome context and their orthologs in the COG database[106]. FunGeCo [107] employs an HMM-based approach to align genes in a newly assembled genome to the Pfam database[108] and records their location information. Based on this information, FunGeCo infers the significant occurrence of domains in gene context and visualizes them on a web server. Further, FlaGs[109] extracts the upstream and downstream genes for the given gene of interest, annotates and further clusters the flanking genes using a sensitive HMM-based method. Then a phylogenetic tree is further constructed for those flanking genes to visualize their conservation.

### 3.3. MAG taxonomic profilers

Identification and quantification of microbial taxa present within the MAGs is an important area for metagenomics studies. The taxonomic classification and relative abundance of the sampled microbial organisms comprising the MAGs are essential to be characterized. To achieve this, one of the most common approaches is to align the reads/contigs to MAGs or reference genomes. However, the alignments are relatively slow to thousands of MAGs and the available reference genomes are usually incomplete. Taxonomic metagenome profilers are used to predict the relative abundances of microorganisms and taxonomic identities of the sampled microbial organisms within the complex community. In contrast to taxonomic binning, the taxonomic profiling method does not assign any individual sequences but offers a snapshot of the presence and relative abundance of diverse taxa within the microbial community[110].

**Tools for MAG taxonomic classification**

The methods based on 16S ribosomal RNA small subunit genes have been successfully established to catalogue and understand the diversity of MAGs in prokaryotic communities, but they offer limited resolution and 16S ribosomal RNAs are poorly representations of whole bacterial genomes. In contrast, the methods based on the single-copy marker genes have gained popularity due to their improved resolution (Table 6). GTDB-Tk[111] adopts HMMER[112] to identify the marker genes in the genomes from the comprehensive Genome Taxonomy Database[113]. These marker genes are concatenated for multiple sequence alignment and then used to construct a reference tree by a likelihood-based phylogenetic inference algorithm[114]. GTDB-Tk performs taxonomic classification for a query MAG based on its position in the reference tree, relative evolutionary divergence, and average nucleotide identity to the reference genomes. ezTree[115] is designed to automatically search for single-copy marker genes for the sequences in the database and builds phylogenetic trees based on maximum-likelihood[116]. To improve the classification of closely related MAGs, Asnicar F *et al.* develops PhyloPhlAn 3.0 [117], which has extracted species-specific marker genes from the integration of >150,000 MAGs and >80,000 reference genomes. PhyloPhlAn designs a novel strategy to place the query MAGs in the phylogenetic tree. The steps include identifying the marker genes from the query MAG sequences, multiple sequence alignment (MSA) for the marker genes, concatenation of their MSAs into a unique one, reconstruction and refinement of the phylogeny using a maximum-likelihood approach. In addition, Microbial Genomes Atlas[118] uses a unique method to classify query MAGs by evaluating the similarities between the query MAGs and the sequences from the reference database, which are calculated based on the genome-aggregate average nucleotides and amino acid identities. The reference sequence with the best matching score will be selected using the Markov clustering algorithm.

**Tools for profiling MAG abundance**

Most existing tools require the microbial reference genomes to extract the genome characteristics or to allow the abundance calculation. The availability of a huge number of MAGs provides an unprecedented opportunity to largely extend the reference database and improve our understanding of the distributions of microbial composition and abundance. There are a few tools available for the tasks, and they have been categorized into several categories, such as protein-based tools such as Kaiju[119], k-mers-based tools such as Kraken[120], k-SLAM[121], and CLARK[122], and marker gene-based tools such as MetaPhlAn2[123] and IGGsearch[124], and SNP-based tools such as ConStrain[125], Strain Finder[126], and Strainest[127]. All these tools are MAG abundance profilers, but they are developed to perform distinct functions in profiling MAGs. For example, k-mers-based tools calculate the abundance of the specific sequences of MAGs whereas the marker gene-based tools report their taxonomic abundance. Here we have discussed those tools and their potential roles in MAG profiling (Table 7).

**Protein-based tools**

The most commonly used protein-based MAG abundance profiler is Kaiju[119]. It is a protein-level classification tool dedicated to the sensitive taxonomic classification of a large number of reads from metagenomic or metatranscriptomic datasets. As the first step, it compacts the reference amino acid sequences, such as proteins predicted from MAGs, by Burrows-Wheeler transformation (BWT)[119] and arranges each sequence by FM-index to reduce computational time and memory cost. Next, Kaiju translates the query sequence into amino acids, aligns against the reference protein database and sorts all the resulting alignments. Once Kaiju detects any sequences homology in the reference database, it usually outputs the taxonomic identifier of the best match. And sometimes it also determines the lowest-common ancestor (LCA) upon recognizing the substantially good matches among the different taxa. As Kaiju employs protein-level classifications, it ensures greater sensitivity against the methods relying on nucleotides. Kaiju usually utilize the available complete genomes from RefSeq or microbial subset of non-redundant (nr) protein database of NCBI BLAST, which can be easily extended to MAGs. It can extend its search capacity to the fungi and sometimes to the microbial eukaryotes depending on the context. Within Kaiju, the reads are translated into amino acid sequences, and then searched against the database to recognize the maximum exact matches.

**k-mer based tools**

Kraken, the most popular k-mer based tool, replaces sequence alignment with searching on a simple k-mer lookup table[14]. In Kraken, the k-mers from the built-in database are saved in a compressed lookup table that can be promptly queried for exact matches to k-mers found in the reads. For each query read, a tree is constructed using the k-mers from the read and associated taxa's ancestors and then used to determine the final classification with maximal root-to-leaf path. Compared with Kraken, Kraken2[128] maintains the accuracy but reduces the memory and computational requirements, enabling more reference genomes in the database. Bracken[129], an extension of Kraken, uses a Bayesian probability model to estimate the abundance of microbial species. Another tool, CLARK[122] uses the discriminative k-mers to perform a supervised sequence classification and reports the assignments with confidence scores[122]. k-SLAM[121], another k-mer based approach that uses local sequence alignments and pseudo-assembly strategies to generate contigs leading to more specific assignments of taxonomic classifications[121]. The taxonomy inference is then performed with the lowest common ancestor technique on the taxonomy tree.

**Marker gene-based tools**

MetaPhlAn2[123] is extensively used among the tools for metagenomic profiling based on marker genes. For any given sample, MetaPhlAn2 aligns the reads to a built-in marker gene database and normalized the counts on each gene. From such genome-scale information, the abundance of each species is estimated from the average abundance of genes consisting of those species. Based on the definition of "dominant strain" per species, StrainPhlAn[130] uses the same marker gene set to construct species' consensus sequence and infers the strain-level genotypes. By contrast, PanPhlAn[131] maps the reads against the species' pangenome and provides the presence/absence information at the gene level. In addition, another tool, IGGsearch[124] adopts a similar approach with MetaPhlAn2, but it extracts the pools of marker genes from the reconstructed MAGs that were constructed from 3,810 fecal metagenomes[124]. It stretches the boundary of the metagenomics field to explore the phylogenetic diversity of different bacteria and other prokaryotes. However, IGGsearch is currently limited to be exclusively deployed within the human gut microbiota.

**SNP-based tools**

ConStrain[125] is the most prominent SNP-based tool to identify the strain-level genotypes. It mainly applies the SNP-flow and SNP-type clustering algorithms on SNP profiles in universal genes to identify the mixture of strains. The relative abundance of each strain is further measured with the Metropolis-Hastings Markov Chain Monte-Carlo approach. Strain Finder[126] postulates a multinomial distribution model for observed SNPs at a given position to identify multiple strains in a species. Then it uses the expectation-maximization algorithm to maximize the likelihood to measure the strain frequencies and their genotypes. By contrast, StrainEst[127] adopts a penalized optimization procedure to detect all the strains within a species of interest.

**4. Outlook, potential challenges, and strategies to address them**

The genome assembly from metagenomic sequencing data is an essential step to produce MAGs. Different assemblers have been developed (as discussed in section 2.1) to perform this. They are ideally required to generate all the high, medium, and low abundance microbial genomes. However, they mostly fail to distinguish the reads from low abundance microbes and the contamination in library preparation and sequencing. Even we can generate MAGs with low abundance species, commonly the contig contiguity is relatively poor. Unfortunately, the contigs generated by short-read assemblers are incomplete genomes, and thus, they still require binning tools to group them as MAGs. On the other hand, long-read sequencing may be a promising technology to produce complete genomes[132]. However, concern remains with the long-read sequencing technologies. It is prone to consist higher sequencing errors, thus, the circularization process could be a challenge sometimes. At this point, it demands some efficient algorithms to be developed, especially for the complete and plasmid genome assembly, where it is mostly limited.

Contig binning is another critical step in generating MAGs. Most of the available binning tools require long contigs (at least >1Kb) as they require sufficient nucleotides to estimate the TNF and read depth. In our previous study[66], we identified that most contigs were below this required threshold, thus could not be clustered using most of the existing tools. The graph-based tools have been recently designed to solve this issue by considering the links between long and short contigs based on sequence overlap and paired-end constraints[66],[133],[134]. They have certainly improved the performance, but more investigations should be performed in this area, especially while introducing the assembly graphs from higher error-prone long-reads. The contigs may be derived from the genomic sequences those could be either homologous sequences or obtained from horizontal gene transfer. These contigs should be carefully assigned into multiple bins, but most current tools simply assign them to the most likely bins or keep them unbinned. Recently, GraphBin2[135] has been proposed to allow the contigs belonging to multiple bins by considering contig connectivity and read depth[135]. The feature space for contig binning is usually high due to a large number of TNF, and at the same time, most of the clustering algorithms fail to achieve good performance on the high dimensional data. A recent study proposes a deep learning model VAMB[67] to apply variational autoencoders to represent the contigs in low dimensions, which significantly reduces the hidden noise and makes the clustering more accurate. Certainly, in the near future, more sophisticated deep learning models will emerge in this field.

Due to computational and experimental limitations, gene families with irregular features and with short sequence lengths are usually neglected in gene prediction. But they may play potentially substantial roles and have critical biological functions relating to host-microbe interaction, defense against phage and bacterial adaptation[136]. The recent advancements of deep learning-based models for gene prediction without requiring manual feature selection certainly promise to improve prediction efficiency and quality[34,59,84]. Next, lacking a ubiquitously accepted tool for strain-level characterization hinders the huge efforts in microbial identification. Perhaps, this happens due to the inherent heterogeneous populations of strains, including low abundance and extremely closely related strains. In this case, long-read sequencing may be promising to distinguish the strains easily by considering the variant phasing[43,48]. To confront such issues, the single-cell metagenomic sequencing perhaps could play wisely, as it isolates the microbial single cells and sequences them directly, which could significantly improve the MAG qualities by deconvolving the mixture of species[12,15]. Despite having some limitations[7,22], the field of constructing MAGs from metagenomic data is rapidly evolving and constantly gains immense complements in decoding the cryptic features of the microbiome. Many different types of methods have been developed to deal with the dedicated issues of producing MAGs, but there are still gaps that need to pay much attention to in future development.

Metagenomics has been evolved immensely in the last decades, but a certain extent of technical challenges and pragmatic constrictions deter the ability to isolate and sequencing every single constituent species within the microbiota as well as the microbiome. Despite having a disparity in their definitions, they are quite interchangeable concepts in metagenomics[137,138]. On the other side, studying metagenomics offers access to the huge uncultured microbial diversity. However, in most cases, ~50% of those putative species remain uncultured and are not merely classified genus-wise. With the advancements of metagenomic tools and assembly tools, better MAGs are reconstructed. In a further extension, Almeida *et al*. demonstrated a representative genome of 92,143 MAGs reconstructed from human gut assemblies[139]. It was able to categorize 73% of the underlying read data. Besides, having the more comprehensive reference genome of the microbiome must ensure better MAGs that can facilitate exploring deep insights of diverse species and their influence within the complex microbial community. The advancement of metagenomic approaches help to steer the microbiome to move towards mechanisms and then to decode many mysterious interactions among the host genome and microbiome[4,140]. Both culture-dependent and culture-independent *de novo* assemblers are essentially inclined towards the most abundant organisms, thus the less abundant species may still be missed[29]. However, having access to the pervasive collections of bacterial genomes from pure culture to MAGs certainly offers the ability to perform efficient and precise reference-based genome analyses to accomplish a thorough classification of complex microbial communities[29].

In many situations, the current status of MAGs is still inadequate, and several challenges have been identified – including low recovery rates for some species, lower species abundance, higher diversity in strains[9]. Despite having an enormous effort to culture and sequence the microbial species of the gut microbiome, many of them refuse to be cultivated within the lab environment. It is a genuine concern at this time because, on one side, they are prevalent in the human gut, and on the other hand, they do not have a well-sequenced genome, thus they are remained out of configuration from the consolidated spectrum genomic characterization. These challenges hinder capturing the holistic views of the microbiome and wide range of species identification. Possibilities of species identification and assessing their influence are highly correlated with the quality of assembly studies. The large-scale metagenomic assembly and binning approaches are quite efficient in recovering the genomes for previously undetermined microbiome species [9]. Thus, by leveraging the power of metagenomics, the unprecedented power of microbiome influencing human health and diseases could be decoded from characterization to mechanistic insights.

**CRediT authorship contribution statement:**

**Chao Yang:** Writing - original draft, and preparing tables; **Debajyoti Chowdhury:** Writing - original draft, and visualizations; **Zhenmiao Zhang:** Consolidating resources; **William K. Cheung:** Supervision; **Aiping Lu:** Supervision; **Zhao Xiang Bian:** Supervision; **Lu Zhang:** Project administer, Writing – review & editing, Supervision, and funding acquisition.

**Declaration of Competing Interest:**

The authors declare that they have no known competing financial interests or personal relationships that could have appeared to influence the work reported in this paper.

**Acknowledgments:**

L.Z. is supported by a Research Grant Council Early Career Scheme (HKBU 22201419), an IRCMS HKBU (No. IRCMS/19-20/D02), an HKBU Start-up Grant Tier 2 (RC-SGT2/19-20/SCI/007), two grants from the Guangdong Basic and Applied Basic Research Foundation (No. 2019A1515011046 and No. 2021A1515012226) and Shenzhen Virtual University Park (SZVUP) Special Fund (2021Szvup135).

**Figure 1:**

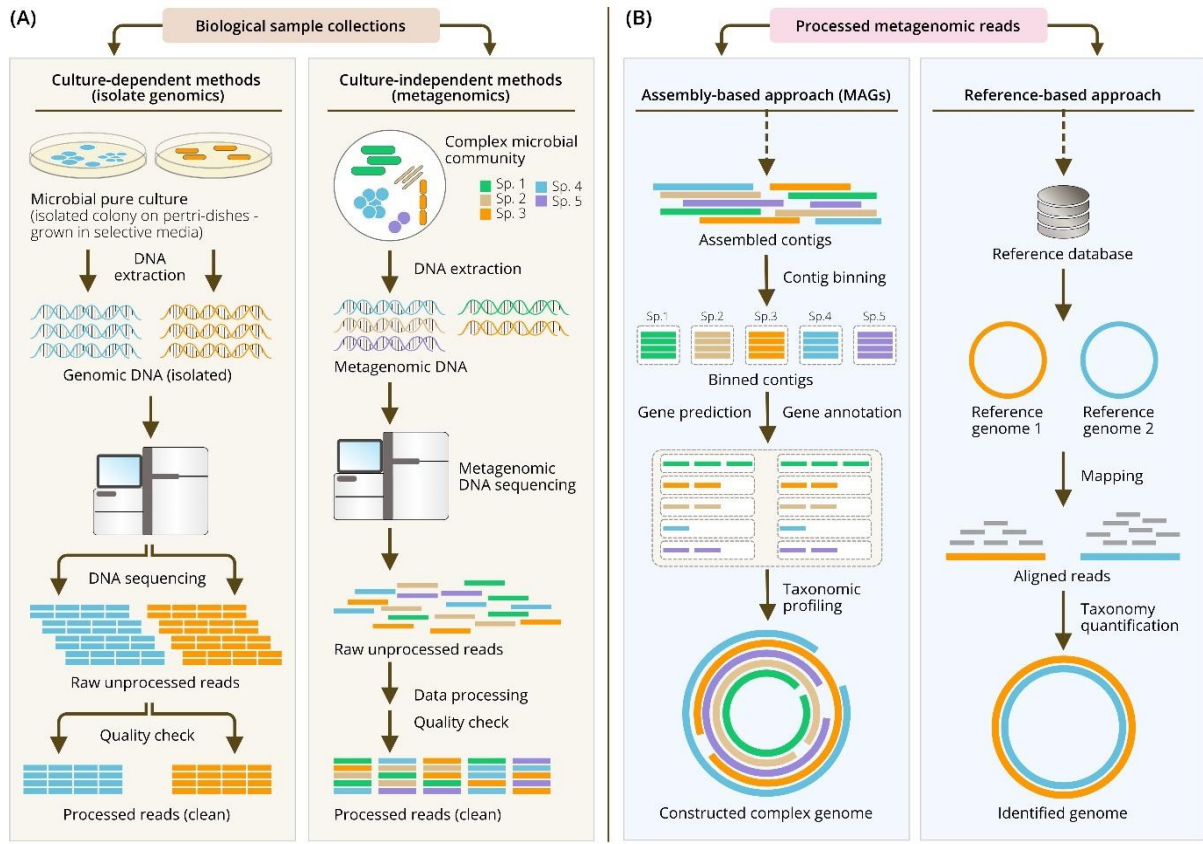

**Figure legends**

**Figure 1: Schematic representation of the different approaches used in the metagenomic research field.**

**(A)** A schematic contrast between culture-independent (metagenomics) approaches, and culture-based genomics approaches: The process for sequencing data generation for the two strategies have been illustrated. **(B)** A schematic contrast between assembly-based and reference-based approaches. The process for step-wise approaches used in assembly-based approaches and reference-based approaches has been illustrated.

**Tables:**

**Table 1: Tools for metagenome assembly**

List of assembly tools sorted in alphabetical order. For each assembler, the adopted technologies (column 2), the original publications (column 3), and summaries of the core algorithms (column 4) and the websites to download these tools (column 5) are illustrated. The assemblers and related algorithms are explained in Section 2.1.

| Tools | Technologies | References | Core algorithms | Websites |
|---|---|---|---|---|
| Athena-meta | linked-reads | Bishara A *et al.* 2018[41] | dbg and local assembly | https://github.com/abishara/athena_meta |
| Canu | long-reads | Koren S *et al.* 2017[47] | OLC | https://github.com/marbl/canu |
| cloudSPAdes | linked-reads | Tolstoganov I *et al.* 2019[42] | dbg | https://github.com/ablab/spades/releases/tag/cloudspades-paper |
| DBG2OLC | short and long reads | Ye C *et al.* 2016[52] | dbg and OLC | https://github.com/yechengxi/dbg2OLC |
| IDBA-UD | short-reads | Peng Y *et al.* 2012[35] | dbg | http://www.cs.hku.hk/~alse/idba_ud |
| MEGAHIT | short-reads | Li D *et al.* 2015[36] | dbg | https://github.com/voutcn/megahit |
| metaFlye | long-reads | Kolmogorov M *et al.* 2020[50] | OLC | https://github.com/fenderglass/Flye |
| metaSPAdes | short-reads | Nurk S *et al.* 2017[38] | dbg | https://github.com/ablab/spades |
| MetaVelvet | short-reads | Namiki T *et al.* 2012[31] | dbg | http://metavelvet.dna.bio.keio.ac.jp |
| MetaVelvet-DL | short-reads | Liang KC *et al.* 2021[34] | dbg | http://www.dna.bio.keio.ac.jp/metavelvet-dl/ |
| MetaVelvet-SL | short-reads | Afiahayati *et al.* 2015[33] | dbg | http://metavelvet.dna.bio.keio.ac.jp/MSL.html |
| Nanoscope | linked-reads | Kuleshov V *et al.* 2016[43] | dbg | https://github.com/kuleshov/nanoscope |
| NECAT | long-reads | Chen Y *et al.* 2021[48] | string graph | https://github.com/xiaochuanle/NECAT |
| Omega | short-reads | Haider B *et al.* 2014[30] | OLC | http://omega.omicsbio.org |
| OPERA-MS | short and long reads | Bertrand D *et al.* 2019[53] | dbg and scaffolding | https://github.com/CSB5/OPERA-MS |
| Ray Meta | short-reads | Boisvert S *et al.* 2012[40] | dbg | http://denovoassembler.sf.net. |
| Unicycler | short and long reads | Wick RR *et al.* 2017[54] | dbg | https://github.com/rrwick/Unicycler |
| wtdbg2 | long-reads | Ruan J *et al.* 2020[49] | fuzzy bruijn graph | https://github.com/ruanjue/wtdbg2 |

**Table 2: Tools for assembly quality control**

List of tools for assembly quality control sorted in alphabetical order. For each tool, requires reference genomes or not (column 2), the original publications (column 3) and the websites to download these tools (column 4) are illustrated. The quality control tools and related descriptions are presented in Section 2.2.

| Tools | Require Reference genome | References | Websites |
|---|---|---|---|
| DeepMAsED | No | Mineeva O *et al.* 2020[59] | https://github.com/leylabmpi/DeepMAsED |
| MetaQUAST | Yes | Mikheenko A *et al.* 2016[56] | http://bioinf.spbau.ru/metaquast |
| REAPR | No | Hunt M *et al.* 2013[57] | https://www.sanger.ac.uk/tool/reapr/ |
| VALET | No | Olson ND *et al.* 2019[58] | https://github.com/marbl/VALET |

**Table 3: Tools for Binning**

List of tools for Binning sorted in alphabetical order. For each tool, the adopted technologies (column 2), the original publications (column 3), summaries of the core algorithms (column 4), and the websites to download these tools (column 5) are illustrated. The binning tools and related descriptions are presented in Section 2.3.

| Tools | Technologies | References | Core algorithms | Websites |
|---|---|---|---|---|
| Bin3C | HiC | DeMaere MZ *et al*. 2019[71] | network clustering algorithm | https://github.com/cerebis/bin3C |
| BMC3C | short-reads | Yu G *et al.* 2018[64] | ensemble clustering algorithm | http://mlda.swu.edu.cn/codes.php?name=BMC3C |
| Concoct | short-reads | Alneberg J *et al.* 2014[25] | gaussian mixture models | https://github.com/BinPro/CONCOCT |
| GraphBin | short-reads | Mallawaarachchi V *et al.* 2020[65] | label propagation algorithm | https://github.com/Vini2/GraphBin |
| HiCBin | HiC | Du Yuxuan *et al.* 2021[72] | Leiden algorithm | https://github.com/dyxstat/HiCBin |
| MaxBin2 | short-reads | Wu YW *et al.* 2016[60] | expectation-maximization algorithm | http://sourceforge.net/projects/maxbin/ |
| MetaBAT2 | short-reads | Kang DD *et al.* 2019[61] | label propagation algorithm | https://bitbucket.org/berkeleylab/metabat |
| METAMVGL | short-reads | Zhang Z *et al.* 2021[66] | label propagation algorithm | https://github.com/ZhangZhenmiao/METAMVGL |
| MyCC | short-reads | Lin HH *et al.* 2016[62] | affinity propagation algorithm | http://sourceforge.net/projects/sb2nhri/files/MyCC/ |
| ProxiMeta | HiC | Press M O *et al.* 2017[70] | deconvolution method | https://github.com/phasegenomics/proxiphage_paper |
| SolidBin | short-reads | Wang Z *et al.* 2019[63] | spectral clustering | https://github.com/sufforest/SolidBin |
| VAMB | short-reads | Nissen JN *et al.* 2021[67] | deep variational autoencoders | https://github.com/RasmussenLab/vamb |

**Table 4: Tools for gene prediction**

List of tools for gene prediction sorted in alphabetical order. For each tool, the method classifications (column 2), the original publications (column 3), summaries of the core algorithms (column 4) and the websites to download these tools (column 5) are illustrated. The gene prediction tools and related descriptions are presented in Section 3.1.

| Tools | Types | References | Core algorithms | Websites |
|---|---|---|---|---|
| Balrog | machine learning | Sommer MJ *et al.* 2021[86] | convolutional neural network | https://github.com/salzberg-lab/Balrog |
| CNN-MGP | machine learning | Interdiscip Sci *et al.* 2019[85] | convolutional neural network | https://github.com/rachidelfermi/cnn-mgp |
| FragGeneScan | model based | Delcher *et al.* 2007[80] | hidden Markov model | https://omics.informatics.indiana.edu/FragGeneScan/ |
| Glimmer-MG | model based | Kelley DR *et al.* 2012[79] | interpolated Markov model | https://github.com/davek44/Glimmer-MG |
| MetaGene | model based | Noguchi *et al.* 2006[82] | dynamic programming | http://metagene.nig.ac.jp/metagene/metagene.html |
| MetaGeneAnnotator | model based | Noguchi *et al.* 2008[83] | dynamic programming | http://metagene.nig.ac.jp/ |
| MetaGeneMark | model based | Zhu *et al.* 2010[78] | hidden Markov model | http://exon.gatech.edu/meta_gmhmmp.cgi |
| Meta-MFDL | machine learning | Biomed Res *et al.* 2017[84] | deep neural network | https://github.com/nwpu903/Meta-MFDL |
| Prodigal | model based | Hyatt *et al.* 2010[81] | dynamic programming | https://github.com/hyattpd/Prodigal |

**Table 5: Tools for gene annotation**

List of tools for gene annotation sorted in alphabetical order. For each tool, the method classifications (column 2), the original publications (column 3), summaries of the core algorithms/programs (column4) and the websites to download these tools (column 5) are illustrated. The gene annotation tools and related descriptions are presented in Section 3.2.

| Tools | Types | References | Core algorithms | Websites |
|---|---|---|---|---|
| eggNOG-mapper | Homology-based | Huerta-Cepas J et al. 2017[88] | Hidden Markov model | http://eggnog-mapper.embl.de |
| FlaGs | Gene context-based | Chayan Kumar Saha et al. 2021[109] | Jackhmmer (hidden Markov model) | https://github.com/GCA-VH-lab/FlaGs |
| FunGeCo | Gene context based | Anand S et al. 2020[107] | Hidden Markov model | https://web.rniapps.net/fungeco |
| GeConT | Gene context based | Ciria R et al. 2004[105] | Blastp | http://www.ibt.unam.mx/biocomputo/gecont.html |
| GhostKOALA | Homology-based | Kanehisa M et al. 2016[89] | GHOSTX (seed search method) | http://www.kegg.jp/blastkoala/ |
| InterProScan | Motif-based | Quevillon E et al. 2005[102] | Phobius (hidden Markov model) | http://www.ebi.ac.uk/InterProScan/ |
| MG-RAST | Homology-based | Keegan KP et al. 2016[90] | Parallelized BLAT | http://api.metagenomics.anl.gov/api.html |
| PANNZER2 | Homology-based | Törönen P et al. 2018[91] | Sansparallel (suffix array neighborhood search) | http://ekhidna2.biocenter.helsinki.fi/sanspanz/ |

**Table 6: Tools for MAG taxonomic classification**

List of tools for taxonomic classification sorted in alphabetical order. For each tool, the method classifications (column 2), the original publications (column 3), summaries of the core algorithms (column 4) and the websites to download these tools (column 5) are illustrated. The detailed description is presented in Section 3.3.1.

| Tools | Types | References | Core algorithms | Websites |
|---|---|---|---|---|
| ezTree | concatenated protein | Wu YW *et al.* 2018[115] | Maximum likelihood | https://github.com/yuwwu/ezTree |
| GTDBTk | concatenated protein | Chaumeil PA *et al.* 2019[111] | Likelihood-based phylogenetic inference algorithm | https://github.com/Ecogenomics/GTDBTk |
| MiGA | genome-based relatedness | Rodriguez-R LM *et al.* 2018[118] | Markov clustering algorithm | http://microbial-genomes.org/ |
| PhyloPhlAn3 | concatenated protein | Asnicar F *et al.* 2020[117] | Maximum likelihood | https://huttenhower.sph.harvard.edu/phylophlan/ |

**Table 7: Tools for MAG taxonomic profilers**

List of tools for taxonomic identification sorted in alphabetical order. For each tool, the method classifications (column 2), the original publications (column 3), summaries of the core algorithms (column 4) and the websites to download these tools (column 5) are illustrated. The gene prediction tools and related descriptions are presented in Section 3.3.2.

| Tools | Types | References | Core algorithms | Websites |
|---|---|---|---|---|
| Braken | k-mer based | Jennifer Lu et al. 2017[129] | Bayesian probability algorithm | https://ccb.jhu.edu/software/bracken/ |
| CLARK | k-mer based | Ounit R et al. 2015[122] | Spectral decomposition | http://clark.cs.ucr.edu/ |
| ConStrain | SNP based | Luo C et al. 2015[125] | SNP-flow algorithm | https://bitbucket.org/luo-chengwei/constrains |
| IGGsearch | marker gene-based | Nayfach S et al. 2019[124] | Workflow | https://github.com/snayfach/IGGsearch |
| Kaiju | translated protein based | Menzel P et al. 2016[119] | Backwards search algorithm | http://kaiju.binf.ku.dk |
| Kraken | k-mer based | Wood DE et al. 2014[120] | Classification tree algorithm | https://ccb.jhu.edu/software/kraken/ |
| Kraken2 | k-mer based | Wood DE et al. 2019[128] | Spaced seed algorithm | https://ccb.jhu.edu/software/kraken2/ |
| k-SLAM | k-mer based | Ainsworth D et al. 2017[121] | Pseudo-assembly algorithm | https://github.com/aindj/k-SLAM |
| MetaPhlAn2 | marker gene based | Truong DT et al. 2015[123] | Workflow | https://huttenhower.sph.harvard.edu/metaphlan2/ |
| PanPhlAn | marker gene based | Scholz M et al. 2016[131] | Workflow | http://segatalab.cibio.unitn.it/tools/panphlan/ |
| Strain Finder | SNP based | Smillie CS et al. 2018[126] | Expectation-maximization algorithm | https://github.com/cssmillie/StrainFinder |
| StrainEst | SNP based | Albanese D | Penalized | https://github.com/compmetag |
| | | *et al.* 2017[127] | optimization procedure | en/strainest |
| StrainPhlAn | marker gene-based | Truong DT *et al.* 2017[130] | Workflow | http://segatalab.cibio.unitn.it/tools/strainphlan/ |